\documentclass{aa}
\usepackage{graphicx}
\begin{document}
   \title{Giant Planet Formation}

   \subtitle{A First Classification of Isothermal Protoplanetary Equilibria }

   \author{B. Pe\v{c}nik
            \inst{1, 2}
          \and
          G. Wuchterl
          \inst{2}
          }


   \institute{$^{1}$ Max-Planck-Institut f\"ur extraterrestrishe Physik,
              Postfach 1312, 85748 Garching\\
              $^{2}$ Astrophysikalisches Institut und Universit\"{a}ts- Sternwarte,
              Schillerg\"{a}$\mathrm \ss$chen 2-3, 07745 Jena
               \\
              \email{bonnie@astro.uni-jena.de}
              }

   \date{Received: May 6, 2003; Accepted: December 17, 2004}

   \abstract{We present a model for the equilibrium of solid planetary cores embedded in a gaseous nebula.
   From this model we are able to extract an idealized roadmap of all hydrostatic states of the isothermal
   protoplanets. The complete classification of the isothermal protoplanetary equilibria should improve the
   understanding of the general problem of giant planet formation, within the framework of the nucleated instability
   hypothesis. We approximate the protoplanet as a spherically symmetric, isothermal, self-gravitating classical ideal gas envelope in
equilibrium, around a rigid body of given mass and density, with
the gaseous envelope required to fill the Hill-sphere. Starting
only with a core of given mass and an envelope gas density at the
core surface, the equilibria are calculated without prescribing
the total protoplanetary mass or nebula density. In this way, a
variety of hydrostatic core-envelope equilibria has been obtained.
Two types of envelope equilibria can be distinguished:
\emph{uniform} equilibrium, were the density of the envelope gas
drops approximately an order of magnitude as the radial distance
increases to the outer boundary, and \emph{compact} equilibrium,
having a small but very dense gas layer wrapped around the core
and very low, exponentially decreasing gas density further out.
The effect of the envelope mass on the planetary gravitational
potential further discriminates the models into the
self-gravitating and the non-self gravitating ones. The static
critical core masses of the protoplanets for the typical orbits of
1, 5.2, and 30 AU, around a parent star of 1 solar mass
($\rm{M_\odot}$) are found to be 0.1524, 0.0948, and 0.0335 Earth
masses ($\rm{M_\oplus}$), respectively, for standard nebula
conditions (Kusaka et al. \cite{kusaka}). These values are much
lower than currently admitted ones primarily because our model is
isothermal and the envelope is in thermal equilibrium with the
nebula. Our solutions show a wide range of possible envelopes. For
a given core, multiple solutions (at least two) are found to fit
into the same nebula. Some of those solutions posses equal
envelope mass. This variety is a consequence of the envelope's
self-gravity. We extend the concept of the static critical core
mass to the \emph{local} and \emph{global} critical core mass.
Above the global critical mass, only \emph{compact} solutions
exist. We conclude that the \emph{'global static critical core
mass'} marks the meeting point of all four qualitatively different
envelope regions.
   }

   \maketitle
%

\section{Introduction}

    With the discovery of the extra-solar gas giants, the general
problem of planet formation has considerably grown in complexity
over the last decade. However, a global theoretical overview of
the properties of the giant planets, irrespective of the parent
protoplanetary disc or the total mass of the giant
planet, is still missing.\\
   In the \emph{nucleated instability\/} hypothesis, envelopes of giant planets are thought to be formed as a consequence of
  accretion of solid bodies forming their cores. To determine the envelope mass corresponding to a given core, static
  protoplanetary models have been constructed (e.g. Perri \& Cameron \cite{perri}, Mizuno \cite{mizuno}, Stevenson
  \cite{stevenson}).\\
  If the envelope is modelled including detailed energy transfer
  and if the outer part of the envelope is radiative, and for
  standard assumptions about nebula conditions, it has been found
  that there is an upper limit for the masses of static envelopes
  and therefore for the total mass of a proto giant planet. This
  upper limit in core mass - the critical mass - was found to be
  insensitive to nebula conditions, but to depend weakly on dust
  opacities (Mizuno \cite{mizuno}) and on the rate at which the
  core (solid body) is accreted (Stevenson \cite{stevenson}).\\
  Even the largest static critical masses are typically more than
  a factor of ten smaller than Jupiter's mass (Mizuno
  \cite{mizuno}, Stevenson \cite{stevenson}, Wuchterl
  \cite{wuc91b}, Ikoma et al. \cite{ikoma}). The nondependence of
  the critical mass on nebula conditions disappears when the
  outermost parts of the protoplanetary envelopes become
  convective, which happens for nebula properties which are well
  within of proposed solar nebula conditions (Wuchterl
  \cite{wuc93}). Envelope masses of such protoplanets range between 6 and
  48 Earth masses ($\rm{M_\oplus}$) but hydrostatic models alone
  are unable to reproduce a Jupiter-mass protoplanet. Therefore
  dynamical and/or quasi-hydrostatical effects should
  play an important role in the formation of gas giants.\\
  There is a number of incompletely studied processes (e.g. the
  formation, evolution, and stability of the protoplanetary disks,
  the dust growth, the planetesimal formation, etc. ) that are
  relevant for the general problem of planet formation. Their
  complexity makes a piecewise approach necessary in studies of
  planet formation. An alternative approach is to study the
  final outcome, i.e. the possible and probable end-states of the
  process. In that context, we present an idealized road-map of
  all hydrostatic states, in order to provide insight when
  analyzing the complex behavior of hydrodynamic and
  quasi-hydrostatic models with detailed microphysics. In
  addition, this work aims to clarify the concept of the critical
  core mass necessary to permanently attract gas of the
  protoplanetary nebula to a terrestrial-planet-like heavy
  element core.

\section{Model}

\subsection{Motivation}

Within nucleated instability theory, the formation of giant
planets includes many possible scenarios for protoplanetary cores
and their respective envelopes. These range from small planetoids
embedded in dilute protoplanetary nebulae to
present-day-Jovian-like cores of several $\rm{M_\oplus}$ squeezed
by some Mbars of metallic H$_2$-He mixtures (Guillot
\cite{guillot}). To date, many investigations have been made into
the evolution of protoplanets, both hydrostatically (e.g.
Bodenheimer et al. \cite{bodenheimer}, Ikoma \cite{ikoma}, see
Wuchterl et al. \cite{ppiv} for review) and hydrodynamically (e.g.
Wuchterl \cite{wuc91a}, \cite{wuc91b}, \cite{wuc93}). In these
studies, 'the evolution' of \emph{particular} planets is followed,
but not much is known about the evolution of all possible
protoplanets. Therefore, it is somewhat \emph{difficult to bring
the detailed solutions of previous investigations within a global perspective}.\\
We follow the thermodynamical approach that was used by Stahl et
al. (\cite{stahl}) to investigate the coreless equilibria of
constant-mass isothermal gas spheres, and the nature of the Jeans
instability. We also expand on the work of Sasaki \cite{sasaki},
who studied isothermal protoplanets in the minimum mass solar
nebula (MMSN). In our model the total mass of the protoplanet and
the density of nebula cloud in which the protoplanet is embedded
are not prescribed. In leaving these as output variables, and
starting only with the (heavy-element) core mass and the density
of the envelope gas at the core's surface, we aim for a
\emph{complete classification of all hydrostatic equilibria}. This
classification should contribute to clarifying whether multiple
planetary equilibria exist for given nebula conditions and how
protoplanetary models relate to gas giants, both inside and
outside of the solar system.

\subsection{Model Assumptions}

We approximate the protoplanet as a spherically symmetric, isothermal, self-gravitating classical ideal gas
envelope in equilibrium around a core of given mass. This gaseous envelope is that required to fill the
gravitational sphere of influence, approximated by the Hill-sphere:
\begin{equation}\label{r_Hill_gen}
  r_{\rm Hill}=a\sqrt[3]{M_{\rm planet}/3M_{\star}},
\end{equation}
where $a$ is the orbital distance from a parent star. With mean
molecular weight of $\mu=2.3~10^{-3} \; \rm{kg\,mol^{-1}}$,
protoplanetary envelopes, as well as the nebula, are roughly
approximated by a hydrogen-helium mixture. The protoplanet's
heavy-element-core is represented by a rigid sphere of uniform
density of $\varrho_{\rm core}=5500 \rm {\, kg \, m^{-3}}$.

The nebula temperature profile is taken according to Kusaka et al.
(\cite{kusaka}), and Hayashi et al. (\cite{hayashi}), cf. Table
\ref{TabManifolds}. The nebula density structure is not \emph{a
priori} determined, but, for critical core mass determination,
nebula densities agree with those from Kusaka et al.
(\cite{kusaka}) for $a=1$ and 30 AU, and from Hayashi
(\cite{hayashi}) for $a=5.2$ AU, cf. Table \ref{TabManifolds}. It
has been shown that the critical core mass values have only a weak
dependence on the nebula density (cf. Sect.~\ref{SectStatCCM}),
therefore the choice of the nebula density is not critical.

\subsection{Model Equations}

The envelope is set in isothermal hydrostatic equilibrium, with
spherical symmetry, and as such is described by:\\
\begin{equation}\label{poisson}
  \frac{\textit{d}M(r)}{\textit{d}r}=4\pi r^2\varrho(r),
\end{equation}
the equation of hydrostatic equilibrium:
\begin{equation}\label{hyd_equil}
\frac{\textit{d}P(r)}{\textit{d}r}=-\frac{GM(r)}{r^2}\varrho(r),
\end{equation}
and the equation of state for an ideal gas:
\begin{equation}\label{id_gas}
  P(r)=\frac{\Re T}{\mu}\varrho(r).
\end{equation}
$M(r)$ is defined as the total mass (core plus envelope) contained
within the radius $r$:
\begin{equation}\label{mass_radial}
  M(r)=M_{\rm core}+\int_{r_{\rm core}}^r4\pi r'^2\varrho(r')\,dr',
\end{equation}
where $r$ is the radial distance measured from the core center and
$\varrho$ is the envelope gas density at radial distance $r$.

\subsection{Boundary Conditions}

The total mass of the protoplanet is defined as:
\begin{equation}\label{tot_mass}
  M_{\rm tot}=M_{\rm core}+M_{\rm env}=M(r_{\rm out})
\end{equation}
with
\begin{equation}\label{M_core}
  M(r_{\rm core})=M_{\rm core}.
\end{equation}
The inner and outer radial boundaries are:
\begin{equation}\label{r_core}
  r_{\rm in}=r_{\rm {core}}=\sqrt[3]{\frac{M_{\rm core}}{\frac{4}{3}\pi\varrho_{\rm
  core}}} \;\;\;\; {\rm{and}} \;\;\; r_{\rm out}=r_{\rm Hill}.
\end{equation}
An additional boundary condition at the core surface is:
\begin{equation}\label{rho_core}
  \varrho_{\rm env}(r_{\rm core})=\varrho_{\rm csg}.
\end{equation}

This model, together with the specified assumptions and boundary conditions, is sufficient to completely determine
a single model-protoplanet. The total mass and nebula density at $r_{\rm Hill}$ (gas density at protoplanet's
outer boundary) are results of the calculation.

   \begin{figure*}
   \centering
   \includegraphics[width=18cm]{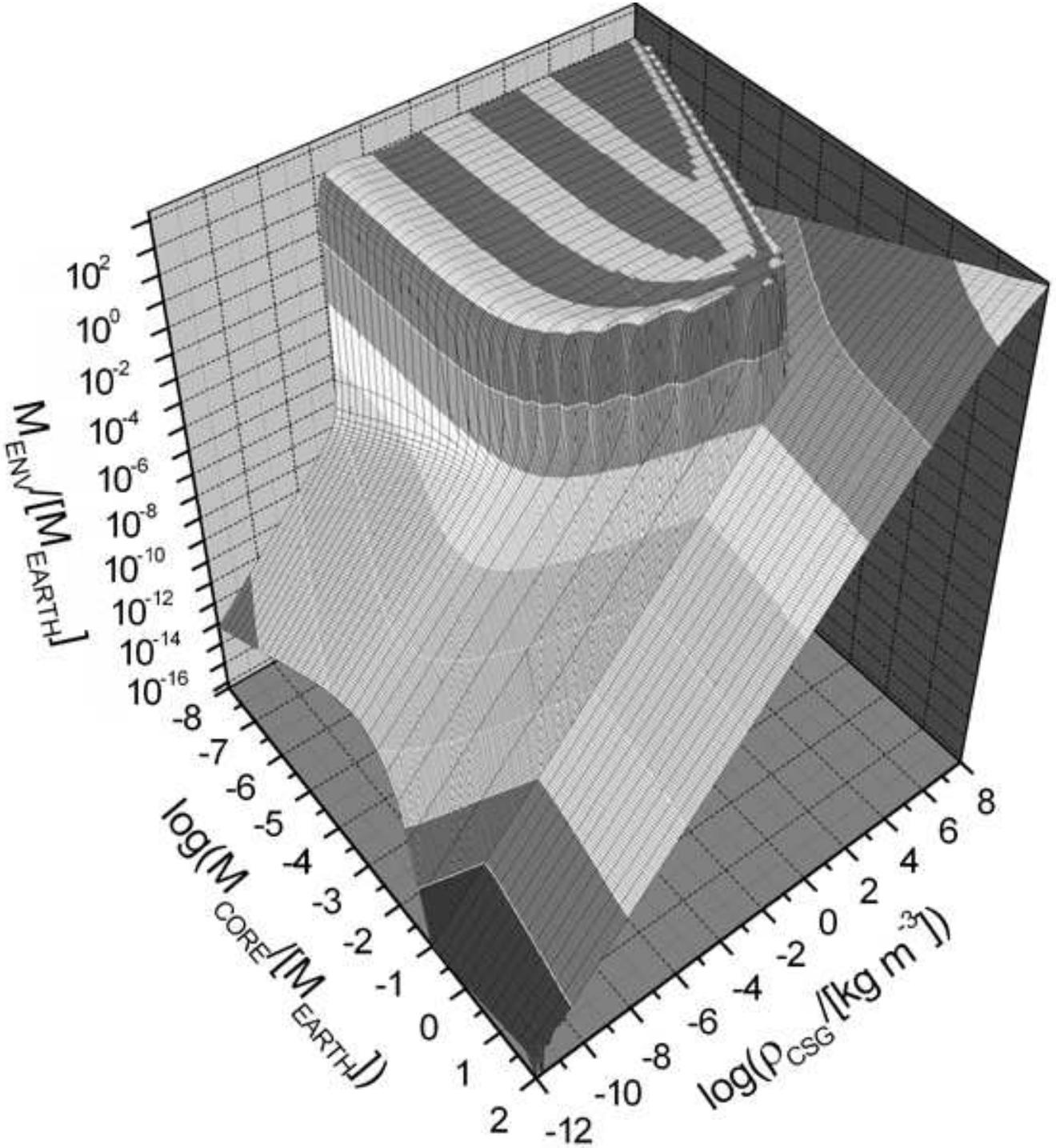}
      \caption{Envelope mass solution manifold. Environmental parameters for this manifold are set to
$a$=5.2 AU, and $T$=123 K. Each point on the surface gives the
mass of the protoplanet's envelope for a given $M_{\mathrm
{core}}$ and gas density at the core surface, $\varrho_{\mathrm
{csg}}$. Models with different initial parameters generally
connect to different nebulae. Several different regions are easily
discernible: I - flat slope with gradient of 1, for the region
[-1,2] in $\log\rm{M_{core}}$ and [-12,6] in
$\log\rm{\varrho_{csg}}$; II - flat slope with gradient of 0.5,
roughly encompassing [4-6,8] in $\log\rm{\varrho_{csg}}$, and all
$\log\rm{M_{core}}$; III - 'base of the island', [-8,-1] in
$\log\rm{M_{core}}$ and [-12,-6] in $\log\rm{\varrho_{csg}}$; IV -
'island', [-8,-1] in $\log\rm{M_{core}}$ and [-6,4-8] in
$\log\rm{\varrho_{csg}}$ (cf. Fig.~\ref{FigManifoldReg}). }
         \label{FigMassManifold}
   \end{figure*}

   \begin{table}
      \caption[]{Symbols}
         \label{TabSymbol}
     $$
         \begin{tabular}{l|l}
            \hline
            \noalign{\smallskip}
            $\rm Symbol^{\dagger}$                                    &  Meaning  \\
            \noalign{\smallskip}
            \hline
            \hline
            \noalign{\smallskip}
            $a$ [AU]                                                    & orbital distance  \\
            $G=6.67259 \; 10^{-11}$                                        & gravitational constant\\
            $\mu=2.3 \; 10^{-3}$                                             & mean molecular weight \\
            $M_{\mathrm {core}}$                                        & predefined core mass     \\
            $M_{\mathrm {env}}$                                         & envelope mass     \\
            $M_{\mathrm {tot}}$                                         & total mass \\
            $M(r)$                                                      & total mass interior to radius \textit{r}\\
            $M_{\odot}=1.989 \; 10^{30}$                                   & solar mass\\
            $M_{\oplus}=5.976 \; 10^{24}$                                  & Earth mass\\
            $r_{\mathrm {core}}$                                        & core radius \\
            $r_{\mathrm {Hill}}$                                        & Hill sphere radius \\
            $\Re=8.31441 $                                              & molar gas constant\\
            $\varrho_{\mathrm {core}}=5500 $                            & predefined core density \\
            $\varrho_{\mathrm {csg}}$                                   & envelope gas density at core surface\\
            $\varrho_{\mathrm {env}}$                                   & envelope gas density  \\
            $T(a)$                                                      & nebula gas temperature\\
            \noalign{\smallskip}
            \hline
         \end{tabular}
     $$
     \begin{list}{}{}
        \item[$^{\mathrm{\dagger}}$] SI units used unless otherwise specified
     \end{list}
   \end{table}

\subsection{Solution Procedure}

The total protoplanetary mass is obtained by integrating outward from $r_{\rm core}$ to $r_{\rm Hill}(M_{\rm
tot})$, starting with $r^{0}_{\rm {Hill}}=r_{\rm{Hill}} (M_{\rm {core}})$ and iterating
$r_{\rm {Hill}}(M_{\rm core}+M_{\rm env})$.\\
Integration is performed from the core surface to the Hill radius,
using the Maple 6 software (e.g. Garvan \cite{garvan}), with the
Fehlberg fourth-fifth order Runge-Kutta method.

\section{Results}

\subsection{Solution Manifold}

In order to cover as many hydrostatic solutions as possible, the
system of equations (\ref{poisson}), (\ref{hyd_equil}), and
(\ref{id_gas}) is solved for a wide range of parameters $M_{\rm
core}$ and $\varrho_{\rm csg}$. The set of all solutions for this
range constitutes the solution manifold.
Figure~\ref{FigMassManifold} shows the solution manifold for a
protoplanet whose orbital distance corresponds to the position of
proto-Jupiter according to the Kyoto-model of solar system
formation (Hayashi et al. \cite{hayashi}). The manifolds with
orbital parameters ($a$, $T$) of proto-Neptune and proto-Earth
have similar morphologies. It should be reiterated that \emph{the
solution set contains all qualitatively different protoplanetary
models} at a particular orbital distance; not just for a
particular nebula, but for any nebula, from a dense
gravitationally-just-stable cloud to a near-vacuum space.

   \begin{table}
      \caption[]{Manifolds}
         \label{TabManifolds}
     $$
         \begin{tabular}{l|r|r|r}
            \hline
            \noalign{\smallskip}
            Orb. param. ($a, T$)& (1, 225)     & (5.2, 123)   & (30, 51.1) \\
            \noalign{\smallskip}
            \hline
            \hline
            \noalign{\smallskip}
            $M_{\rm {core,crit}}^{\rm {MMSN}}/[M_{\oplus}]$ & 0.1524 & 0.0948 & 0.0335 \\
            $M_{\rm {env}}^{\rm{max}}/[M_{\oplus}]$ & 21 & 96 & 380 \\
            \noalign{\smallskip}
            \hline
         \end{tabular}
    $$
     \begin{list}{}{}
        \item[] The critical core mass increases for smaller orbital distances because of (in order of importance):
        the higher gas temperature (cf. Sect.~\ref{SectTemp} and \ref{SectT-M}), the smaller Hill sphere
        (cf. Sect.\ref{SectOrbDist}), and the higher densities of the reference nebulae
        (taken from the minimum mass solar nebula models of Kusaka (\cite{kusaka}) and Hayashi (\cite{hayashi})).
     \end{list}
   \end{table}

   \begin{figure}
   \centering
   \includegraphics[width=9.31cm]{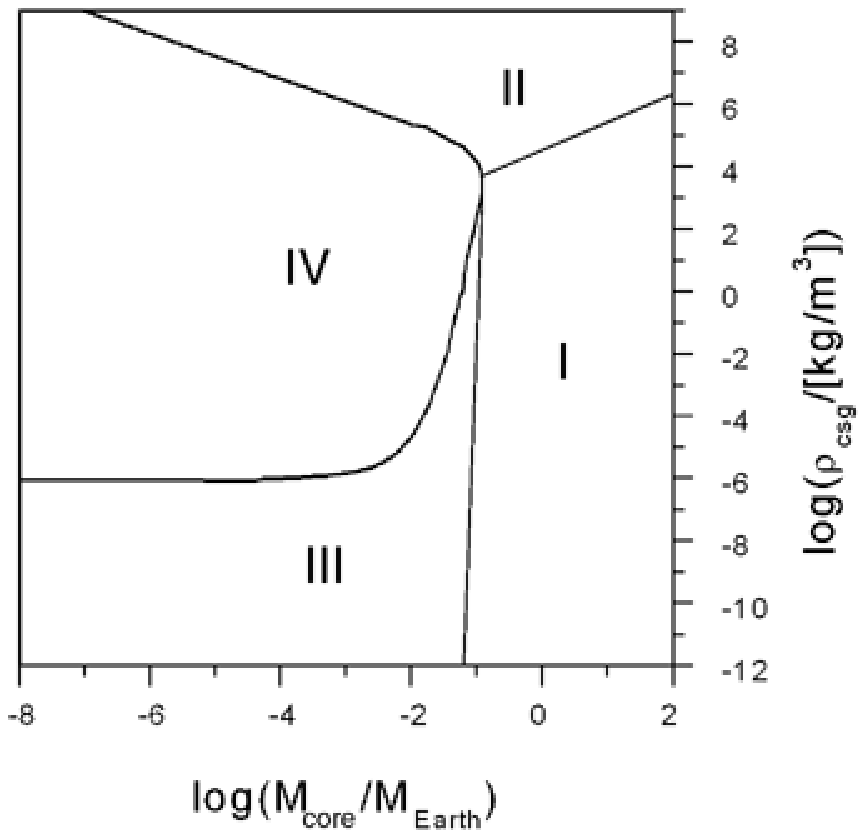}
      \caption{Manifold regions:  I - compact non-self-gravitating
      envelopes, II - compact self-gravitating envelopes, III -
      uniform non-self-gravitating envelopes, IV - uniform self-gravitating envelopes.
      The border of the region IV somewhat depends on the choice of the surrounding nebula
      (cf. Fig~\ref{NebulaeSols}); we use here a value from the Hayashi (\cite{hayashi})
      minimum mass solar nebula model.}
         \label{FigManifoldReg}
   \end{figure}

\subsection{Manifold Regions}
\label{SectManifRegions}

Several distinct regions exist in the parameter space of the
solution manifold (Fig.~\ref{FigManifoldReg}), and they can be
examined from two complementary perspectives. One way is to use
the gas density at the core surface, $\varrho_{\rm{csg}}$, as an
independent variable (e.g. Fig.~\ref{FigSelf_grav}), and the other
is to use the nebula gas density, $\varrho_{\rm{out}}$ (e.g.
Fig.~\ref{FigRho_out_compar}). While $\varrho_{\rm{out}}$ is more
physically intuitive, $\varrho_{\rm{csg}}$ maps out region IV of
Fig.~\ref{FigManifoldReg} more clearly, and is more efficient in
terms of representing the entire manifold.\\
Figure~\ref{FigManifoldReg} divides the solution manifold into
four distinct regions, depending on whether the solution is
compact or uniform and self-gravitating or not.
Figures~\ref{FigSelf_grav} and \ref{FigRho_out_compar} point to
the existence of the four possible regimes for a planet;
\begin{enumerate}
\item \emph{`mature telluric planet'} (region I):
envelope mass is a linear function of $\varrho_{\rm{out}}$, and $\varrho_{\rm{csg}}$
\item \emph{`mature giant planet'} (region II):
envelope mass weakly drops with $\varrho_{\rm{out}}$
($M_{\rm{env}}\propto\varrho_{\rm{out}}^{-0.005}$).\\
$M_{\rm{env}}\propto\varrho_{\rm{csg}}^{0.5}$ is weaker than for
the `nebula' regime. `Nebula' densities ($\varrho_{\rm{out}}$) are
so low that they may well be considered a vacuum.
\item \emph{`nebula'}(region III): envelope mass is a linear function of
$\varrho_{\rm{out}}$ and $\varrho_{\rm{csg}}$
\item \emph{`protoplanet'} (region
IV): envelope mass is a non-trivial function of
$\varrho_{\rm{out}}$ or $\varrho_{\rm{csg}}$
\end{enumerate}

\subsection{Self-Gravity Effect}

\begin{figure}
  \centering
   \includegraphics{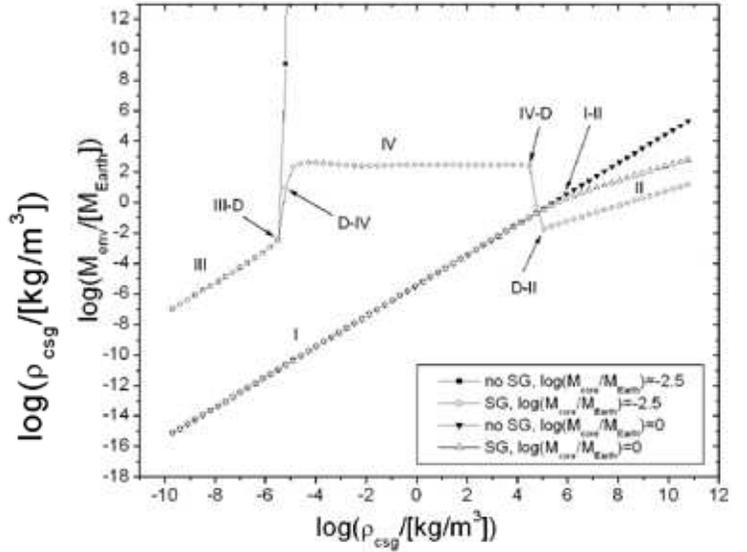}
  \caption{ Demonstration of the self-gravitating effect for sub-
            and super-critical cores: comparison of cuts through
  two manifolds, with- ($M=M(r)$ in Eq.~(\ref{hyd_equil})) and without- ($M=M_{\mathrm {core}}$)
  the envelope's gravitating effect,
  each for two core masses. Cuts are for $a=30$ AU and $T=51.1$ K.
  Circles and squares represent the envelope mass of the subcritical core,
  calculated for $M=M(r)$ and $M=M_{\mathrm {core}}$ in Eq.~(\ref{hyd_equil}), respectively.
  White and black triangles have the same meaning but for the supercritical core.
  Labels without arrows correspond
  to manifold regions from Fig.~\ref{FigManifoldReg}, while labels
  with arrows mark interfaces between regions. $D$ corresponds to
  the 'divergent wall' which surrounds region IV (cf.
  Fig~\ref{FigMassManifold}). Self-gravitating envelopes with $M=M(r)$ in
  Eq.~(\ref{hyd_equil}) have a larger envelope mass than the
  corresponding envelopes with $M=M_{\mathrm {core}}$ in Eq.~(\ref{hyd_equil}) (cf.
  Fig.~\ref{env_types}).} \label{FigSelf_grav}
\end{figure}

\begin{figure}
  \centering
   \includegraphics{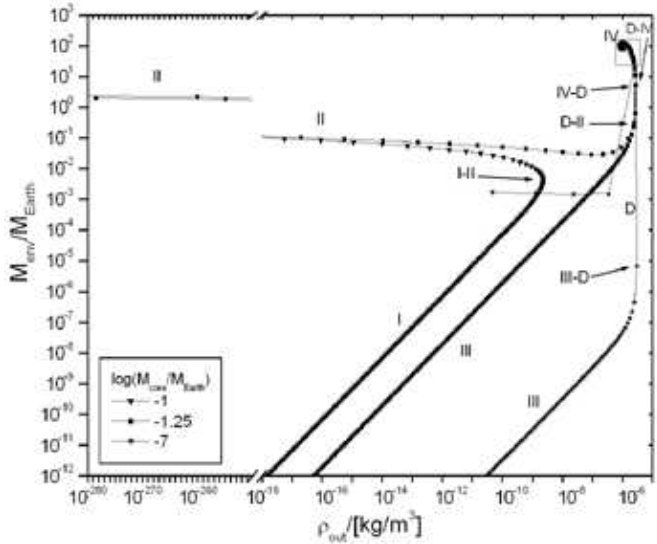}
  \caption{ Envelope mass as a function of the nebula density $\rm{\varrho_{out}}$.
            Labels are the same as in Fig.~\ref{FigSelf_grav}.
            Lines connect states with increasing $\rm{\varrho_{csg}}$.
            Note the strong dependence of $\varrho_{\rm{out}}$ on
            the envelope mass, and a non-trivial behavior of the
            $M_{\rm{env}}(\varrho_{\rm{out}})$ for region IV
            (enlargement in Fig.~\ref{FigRho_out_compar_zoom}).}
  \label{FigRho_out_compar}
\end{figure}

The key effect, which is responsible for the manifold morphology
as observed in Fig.~\ref{FigMassManifold}, can be described as
self-gravity of the protoplanet's envelope. Keeping in mind the
hydrostatics of the model, and the fact that the surrounding
nebula is not prescribed, one can see that \emph{self-gravity
reduces the envelope mass for a given core surface pressure}, i.e.
the envelope mass would be larger if there were no
self-gravitating
effect (Fig.~\ref{FigSelf_grav}).\\
\begin{figure}
  \centering
   \includegraphics{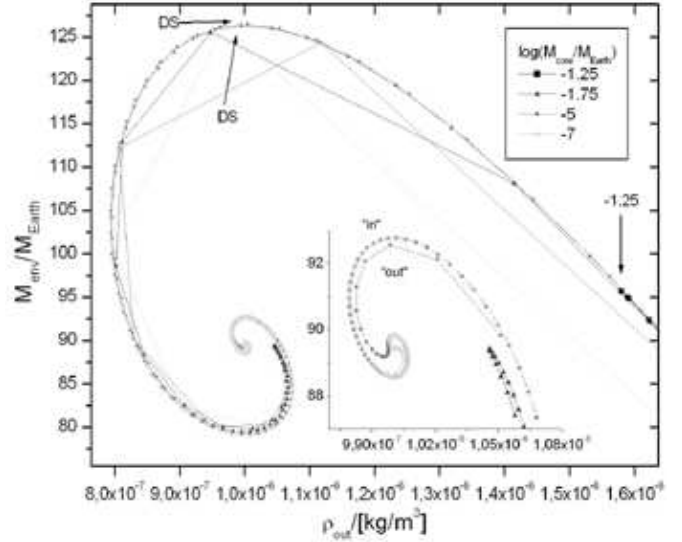}
  \caption{ Enlargement of the boxed region of Fig.~\ref{FigRho_out_compar}, isothermal curl regularized with the
finite-density core; `-1.25': black squares represent protoplanets
with first subcritical $\rm{M_{core}}$ line on the mesh of
Fig.~\ref{FigMassManifold} and the arrow points at the black
square with the highest $M_{\rm{env}}$, $DS$: two protoplanetary
states with the largest envelope mass in the manifold, but with
typically very different $\rm{\varrho_{csg}}$ (cf.
Sect.~\ref{SectDoubleMax}); $in$ and $out$ curves are the
consequence of the core. The smaller the core, the closer the $in$
and $out$ curves are. The figure corresponds to a V-U plane for
the protoplanets (see Sect.~\ref{CorelessVsCore} for further
discussion).}
  \label{FigRho_out_compar_zoom}
\end{figure}
The envelope's radial gas density profile is shaped through the
interplay of inward gravitational force and outward gas pressure.
If the envelope mass is small compared to the core mass, the
gravitational force can be approximated as arising from the core's
gravitational potential only. Once the envelope mass is comparable
to (or greater than) the core mass, they both contribute to the
gravitational potential, making its gradient steeper and, in
effect, reducing the envelope mass. As a consequence, the
self-gravitating envelope connects to a nebula different from the
one which is in balance with the envelope in the absence of the
self-gravitating effect. Further discussion of the role of
self-gravity can be found in Sect.~\ref{CorelessVsCore}.

\subsection{Two types of envelope equilibria}

\begin{figure}
  \centering
   \includegraphics{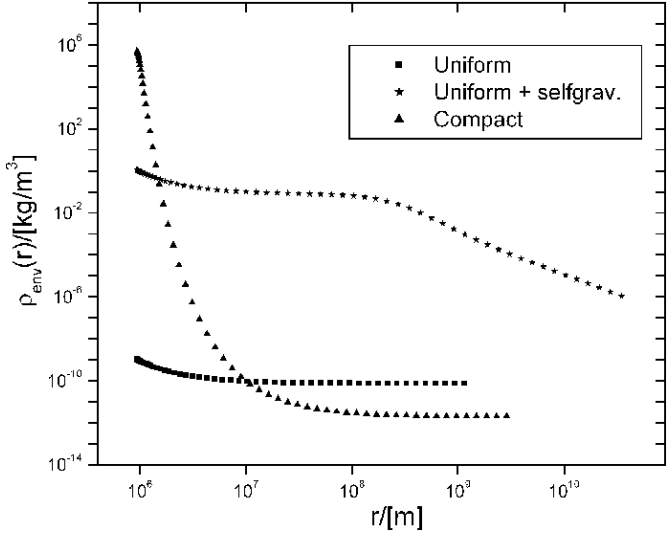}
  \caption{Uniform, compact and self-gravitating profiles. The
  uniform self- gravitating profile resembles the
  non-self-gravitating one until the envelope mass becomes
  comparable to the core mass. Then the density profile
  changes to $\varrho_{\rm env}(r)\propto r^{-2}$.}
  \label{env_types}
\end{figure}

The solution manifold (Fig.~\ref{FigManifoldReg}) contains two
basic types of envelope equilibria (Fig.~\ref{env_types}):
\begin{enumerate}
\item \emph{uniform}, or quasi-homogenous envelope: the density of
the envelope gas drops weakly with increasing radial distance,
keeping the mass distribution more or less uniform throughout the
envelope; $\rm{\partial M_{env}/\partial r_{out}>0}$
\item \emph{condensed}, or quasi-compact envelope: typically small,
but very dense gas layer is wrapped around the core, at larger
radii the gas density is very low; $\rm{\partial M_{env}/\partial
r_{out}\approx0}$
\end{enumerate}
This is reminiscent of a similar equilibrium, found by Stahl et
al. (\cite{stahl}), for constant mass coreless 'Van
der Waals' gas spheres.\\
If an envelope's mass is much smaller than the core mass, the
radial profile of the gas density is simply an exponential
function, well approximated by :
\begin{equation}\label{baro}
  P(r)=P_0\exp(-\frac{\mu}{\Re T} \, GM(r)(\frac{1}{r_{\mathrm{core}}} - \frac{1}{r})).
\end{equation}
If $(M(r)-M_{\mathrm{core}}) \ll M_{\mathrm{core}}$, then Eq.
(\ref{baro}) reduces to the barometric formula.

\subsection{Differences: Isothermal Coreless Gas Spheres vs.
                Isothermal Protoplanets}
\label{CorelessVsCore}

The curl in Figs.~\ref{FigRho_out_compar} and
\ref{FigRho_out_compar_zoom} is reminiscent of a similar feature
found for the isothermal coreless ideal-gas spheres (e.g.
Schwarzschild, \cite{schwarzschild}, Sect.~$13$) represented in
the U-V plane. It follows from the definition of U and V that:
\begin{equation}\label{Eqn_U}
  U=\frac{r}{M(r)}\frac{dM(r)}{dr}=\frac{4\pi r^{3}\varrho}{M(r)}=3\frac{\varrho}{M(r)/(\frac{4}{3}\pi r^{3})}
\end{equation}
\begin{equation}\label{Eqn_V}
  V= -\frac{r}{P}\frac{dP}{dr}=\frac{\varrho}{P}\frac{GM(r)}{r}=\frac{3}{2}\frac{GM(r)/r}{\frac{3}{2}P/\varrho}
\end{equation}
and from the fact that the mean density of the total object for
our model is always the same, as implicitly defined through
Eq.~(\ref{r_Hill_gen}).

Unlike a singular isothermal sphere with an infinite pressure at
the center, our protoplanetary model has a solid core of uniform
(and finite) density at its center. This will result in the
departure from the potential of the coreless isothermal sphere:
instead of having a $\varrho(r)\propto r^{-2}$ structure, the
envelope
gas close to the core surface will obey the form of the barometric law (c.f. Eq.~(\ref{baro})).\\
If the mean envelope density at lower stratifications is
comparable to the core density, an `effective' core will shorten
the characteristic length-scale of the potential, making the
exponential profile of the barometric-law-like profile even
steeper. For the appropriate effective core, the outer
stratifications will exactly match the outer stratifications of
the solution which has the gas density at the core surface much
smaller than the core density (cf. Fig.~\ref{rho_multi}). These
profiles will connect to the same nebula density, but will have a
slightly different envelope mass, because of the difference in the
profile of the inner stratification. Therefore, the curl of
Fig.~\ref{FigRho_out_compar_zoom} will have two branches: `in'
(the solution with a
non-self-gravitating inner stratification) and `out' (the solution with an effective core).\\
The smaller the core mass, the sooner will the profile connect to
the `$r^{-2}$' structure; i.e. the smaller the difference in the
envelope mass between the pairs of solutions, the closer the `in'
and `out' branches in Fig.~\ref{FigRho_out_compar_zoom} will be.

The fall-off of the gas density with increasing radius in the
self-gravitating part of the envelope can be approximated by
$\varrho_{\rm env}(r)\sim r^{-2}$ (cf. Fig. \ref{env_types},
self-gravitating profile), as expected in the theory of stellar
structure for a self-gravitating isothermal sphere of ideal gas
(e.g. Shu~\cite{shu}, Sect. $18$). Small deviations from $r^{-2}$
are due to the finite amount of mass needed for the envelope to
become self-gravitating, which produces a slight imbalance between
the self-gravity and the amount of
mass $M(r)$. No similar effect is observed for coreless, isothermal gas spheres (cf. Stahl et al, \cite{stahl}).\\
Depending on the fraction of the self-gravitating part of the envelope and of the core mass, this wavelike
deviation can extend to the outer boundary, or can be attenuated deep within the envelope.

\subsection{Estimating the Applicability of the Ideal Gas}
\label{SectIdealEst}

We made two major assumptions while constructing our model - that
the gas is ideal, and that the heat is instantaneously radiated
away, i.e. the gas is isothermal. In Sect.~\ref{SectIsoThermEst}
we examine the isothermal assumption, and we deal with the ideal
gas in this section.

In order to keep the protoplanet in an equilibrium with the
surrounding nebula, we have set the envelope gas temperature equal
to the nebula temperature for the appropriate orbital distance.
Therefore, we compare different equations of state at the envelope
temperature. In addition to an ideal gas, we take the
Saumon-Chabrier-van Horn (\cite{SCvH}) EOS, the Carnahan-Starling
(\cite{carnahan-starling}) EOS, as well as a completely degenerate
electron gas.

\begin{figure}
  \centering
   \includegraphics{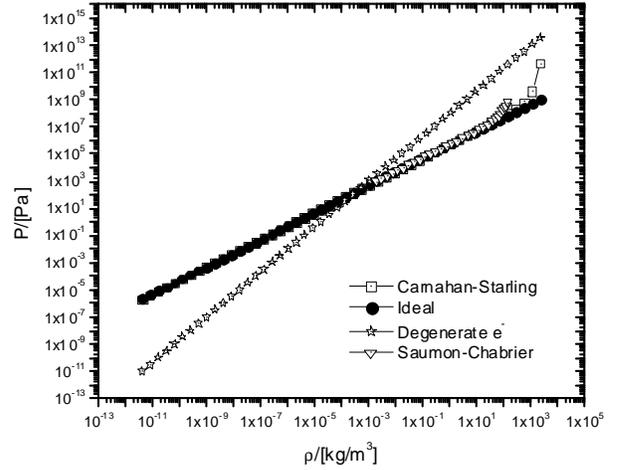}
  \caption {Pressure as a function of density, for $T=123$ K. Black circles represent the ideal gas, squares
  are for the Carnahan-Starling EOS, and triangles are for the Saumon-Chabrier EOS. This figure also shows that
  a completely degenerate electron gas (stars) is not a good assumption for this ($\varrho, \, T$) parameter range.}
  \label{FigEOSCompar}
\end{figure}

Figure~\ref{FigEOSCompar} shows that for the gas densities up to
about $40$~kg\,m$^{-3}$, ideal gas, Saumon-Chabrier-van Horn, and
Carnahan-Starling EOS agree to better than one percent. For higher
densities the Saumon-Chabrier EOS shows additional non-ideal
effects, while the Carnahan-Starling EOS exhibits a similar
behavior for densities larger than $200$~kg\,m$^{-3}$. We can also
see that the electron degeneracy does not contribute to the
pressure at least till the point where the Saumon-Chabrier EOS
departs from ideal-gas behavior.

\begin{figure}
  \centering
   \includegraphics{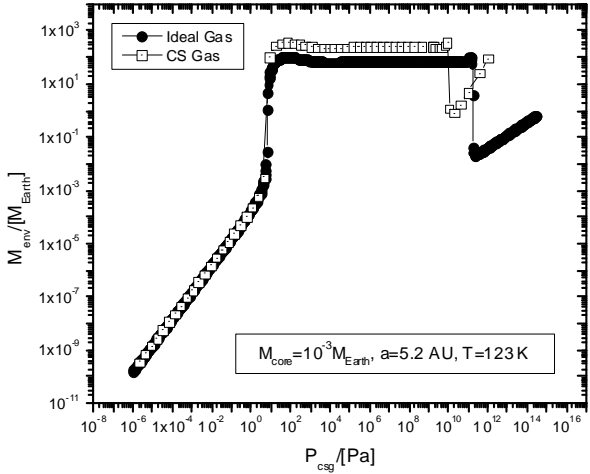}
  \caption {Cut through the envelope mass manifold, for a $10^{-3}M_\otimes$ core, $a=5.2$ AU, and $T=123$ K.
  Black circles represent the ideal gas, and squares are for the Carnahan-Starling EOS.}
  \label{FigMenvEOSCompar}
\end{figure}

However, in general we see that the ideal gas is an excellent
approximation for our model for the better part of the envelope
gas density range. Certainly, there are also models where
densities are high enough for significant non-ideal effects, but
typically for the protoplanets in our model those high density
envelope regions are restricted to areas close to the core, while
the rest of the envelope will be well approximated by an ideal
gas. We can see in Fig.~\ref{FigMenvEOSCompar} that if we use e.g.
the Carnahan-Starling EOS, the numerical details will be changed,
but the qualitative picture will remain the same. This is also
true for the Saumon-Chabrier EOS, which is work in preparation by
C. Broe.g. The ideal isothermal gas will not be a good
approximation for the compact envelopes which are typically
associated with giant planets in the late stages of their
evolution. Using our model, we can show that a protoplanet will
have a compact envelope under certain conditions. What we cannot
do with this model is obtain a quantitatively correct picture of
such a compact envelope.

Additionally, Fig.~\ref{rho_multi} shows why the choice of EOS is
not critical for the qualitative picture: Although the non-ideal
effects might change the density stratifications of the compact
inner parts, each solution which is not self-gravitating in its
inner (barometric-law like) part, will have a counterpart solution
with an effective core. Properties of the effective core will be
dictated by the EOS, but its effect on the scale-height will
remain the same.


\subsection{Estimating the Applicability of the Isothermal Assumption}
\label{SectIsoThermEst}

In the previous section we showed that an ideal gas is a good
approximation for most of the parameter range we use. The validity
of the isothermal assumption is examined below.

\begin{figure}
  \centering
   \includegraphics{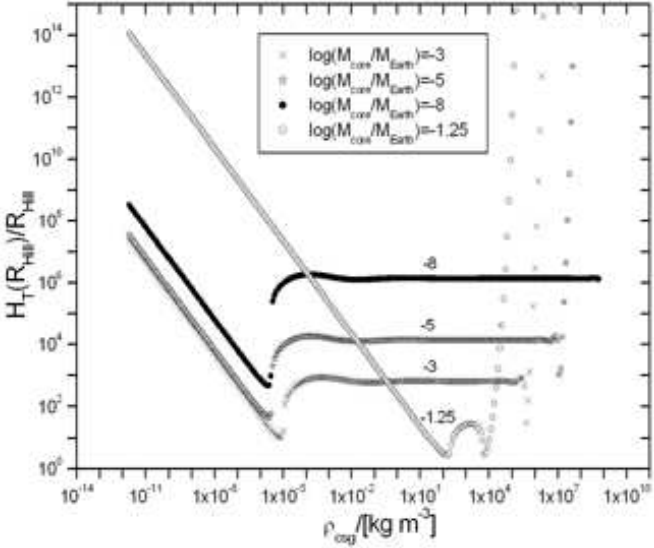}
  \caption {Specific temperature scale-height as a function of the
  density at the core surface, for different subcritical core masses.
  Protoplanetary models with cores of -8 (black circles), -5 (stars), and -3 (crosses)
  in logarithmic $M_\oplus$ units have $\rm{H_T(r_{Hill})}/r_{Hill}$
  much larger than unity. This justifies the isothermal assumption
  for the manifold regions III and IV.}
  \label{FigHTCompar}
\end{figure}

By analogy with the pressure scale-height, a temperature
scale-height of a radiative stratification can be defined as:

    \begin{equation}\label{temp_scale_height}
      H_T = \frac{H_P}{\nabla _{\rm{rad}}}=-\frac{\partial r}{\partial \ln
      T}
    \end{equation}
where

    \begin{equation}\label{pres_scale_height}
      H_P =-\frac{\partial r}{\partial \ln
      P}=\frac{P}{\varrho}\frac{r^2}{G M(r)}
    \end{equation}
for an ideal gas and hydrostatic equilibrium, and
    \begin{equation}\label{nabla_rad}
      \nabla _{\rm{rad}}=\left. -\frac{\partial \ln T}{\partial \ln
      P}\right|
      _{\rm{rad}}=\frac{3\,\kappa\,L\,P}{4\,\pi\,a\,c\,G\,M(r)\,T^4}\;\rm,
    \end{equation}
where $a$ is the radiation constant, $\kappa$ is the gas opacity
taken to be $0.1~\rm{m^2\,kg^{-1}}$, $c$ is the speed of light,
and $L$ is the core luminosity due to the planetesimal accretion
rate of $\rm{10^{-6}~M_\oplus\,yr^{-1}}$.\\
The temperature scale-height corresponds to the length-scale of a radiative giant-protoplanet over which the
envelope temperature drops by a factor of 1/e. The specific temperature scale-height $H_{\rm{T}}(r)/r_{\rm{Hill}}$
evaluates the ratio of the thermal length-scale to the radial extent of the entire protoplanetary envelope, at a
position $r$. Evaluated at $r=r_{\rm{Hill}}$, ${H_{\rm{T}}(r_{\rm{Hill}})}$ is the global estimate of the thermal
scale-height of the protoplanet. Figure~\ref{FigHTCompar} shows that the isothermal assumption is valid for large
portions of the manifold regions III and IV (cf. Fig.~\ref{FigManifoldReg}), where $\rm{H_T(r_{Hill})}/r_{Hill}$
has values much larger than unity. These envelopes have a specific thermal scale-height above unity for at least
the outer 90\% of the envelope. Therefore, even though the small innermost envelope region is probably
non-isothermal, the protoplanet should be well represented by the isothermal gas.\\
Close to the giant-protoplanet's critical core mass (e.g.
$\log\rm{M_{core}=-1.25}$ in Fig.~\ref{FigRho_out_compar}),
$\rm{H_T}/r_{Hill}$ is expected to be of order unity and the
isothermal assumption breaks down. Compact solutions (regions I
and II from Fig.~\ref{FigManifoldReg}, and high $\varrho_{\rm
{csg}}$ solutions in Fig.~\ref{FigHTCompar}) have very large
$\rm{H_T}/r_{Hill}$, indicating that nearly-vacuum space around
the compact envelope is nearly isothermal. Detachment from a
protoplanetary nebula could represent either hydrodynamically
active protoplanets, or the collapsed gas giants with cleared
protoplanetary nebula (i.e. mature giant planets). In both cases
objects are expected to be deep in the non-isothermal regime. The
radial profiles of the compact objects will change if a detailed
energy transport is included, but they will nevertheless remain
compact. A comparison of Jupiter's radius with that of our model
planet (of equivalent mass and $T_{\rm{env}}$ of 5000 K, estimated
to be representative of Jupiter's average temperature from Guillot
\cite{guillot}) shows that, with $r_{\rm{compact}}=6.63\,10^6$ m,
our model falls short by
less than 10\% of reproducing the radius of the real gas giant.\\
In the context of Jupiter's potentially rapid formation (order of
$\rm{10^6}$ years), it could be argued that the core accretion
rate should be even higher. However, $\rm{H_T}/r_{Hill}$ is
proportional to the inverse of $\rm{\dot{M}_{core}}$, and even if
it is set to $\rm{10^{-5}~M_\oplus\,yr^{-1}}$, the validity of the
isothermal assumption is still appropriate for the regions III and
IV of Fig.~\ref{FigManifoldReg}. Indeed, such high core accretion
rates are applicable for cores comparable to $\rm{M_\oplus}$ (i.e.
cores at late stages of a giant protoplanet's evolution), and are
surely an overestimation for the younger cores (e.g. for the cores
of $10^{-3}~\rm{M_\oplus}$), making the case for the isothermal
regime even more solid. However, because of the simplicity of our
model, the results are only qualitative, while quantitatively
correct values would only be accessible through a more elaborate model.\\
$\rm{H_T}/r_{Hill}$ shows that close to the critical core mass there are non-isothermal effects.\\
But the basic isothermal picture is valid for most of (the
quasi-homogenous part of) the manifold. It even appears that the
possible transition from homogenous to compact state can be
initiated within the isothermal regime.

\subsection{Manifolds and Environment}
\label{SectEnviro}

Manifold solutions are dependent on four environmental parameters:
the gas temperature $T$ of the protoplanet (and of the
surrounding-nebula), the orbital distance $a$ from the parent
star, the mean molecular weight $\mu$, and the mass of the parent
star $M_{\star}$. These parameters influence the balance of the
two forces that determine the radial density structure - the
outward force arising from the gas pressure, and the inward force
of gravity; $T$ and $\mu$ are connected with pressure through
Eq.~( \ref{id_gas}), while $a$ and $M_{\star}$
determine the Hill-sphere, i.e. the volume of the envelope mass.\\
Because of the simplicity of the model, the impacts of $T$ and
$\mu$ on the solutions will be discussed together, as will the
influence of $a$ and $M_{\star}$. In reality, these parameters
will have very different impact.

Unless otherwise specified, the reference parameters throughout the current section are: log($M_{\rm
{core}}/M_{\rm {\oplus}}$)=-5, $a=5.2$ AU, $T_{\rm {env}}=123\,\rm K$, and $\mu =2.3~10^{-3} \rm \; kg \; \rm {
mol^{-1}}$.

\begin{figure}
  \centering
   \includegraphics{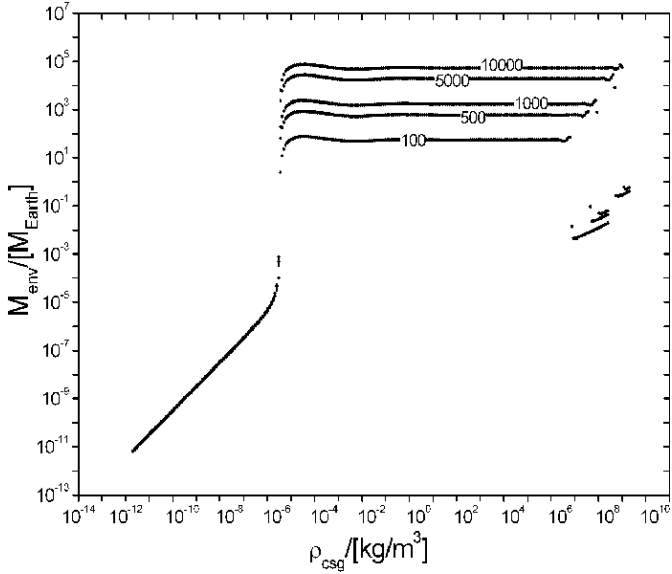}
  \caption {Envelope mass solutions as a function of gas density
            at the core surface, for gas temperatures of 100, 500,
            1000, 5000, and 10000 K. A change of $T$ has no influence
            on the envelope mass of the non-self-gravitating regions,
            while the same change of $T$ will produce a significant effect
            for protoplanets in self-gravitating regions.}
  \label{FigComparManifT}
\end{figure}

\subsubsection{Temperature and Mean Molecular Weight}
\label{SectTemp}

Although \emph{this }model is isothermal, the choice of gas
temperature influences the solution manifold quantitatively. From
Eq.~(\ref{id_gas}) it is clear that pressure relates linearly to
temperature. Since the pressure force counterbalances the
gravitational force, protoplanets with hotter envelopes require
more gravity (and thus more mass) for a hydrostatic solution. The
value of the critical core mass is a good example of the
quantitative influence of the temperature. For example, the
critical core mass for a 123 K protoplanet in Jupiter's
orbit is 0.0948 $M_{\oplus}$, while the critical core mass value for a 5000 K case is 24.5 $M_{\oplus}$.\\
Figure~\ref{FigComparManifT} shows that, for subcritical cores and
low gas densities at the core surface (region III in
Fig.~\ref{FigManifoldReg}), the gas temperature has virtually no
impact on the envelope mass. Since the envelope mass is small
compared to the core mass, the envelope parameters (e.g. $T_{\rm
env}$) have no influence on the hydrostatic force balance via
gravity feedback. On the contrary, for envelopes in which
self-gravity shapes the radial structure (regions IV and II in
Fig.~\ref{FigManifoldReg}), the envelope
mass is significantly affected by different $T_{\rm env}$.\\
The scaling law which relates manifolds of various temperatures is
discussed in Sect.~\ref{SectT-M}.\\
As previously mentioned, this simple model does not incorporate an
energy transport equation, nor does it take into account the gas
and dust opacities. Therefore, a change in $\mu$ cannot be
distinguished from the corresponding change in $T$, and will not
be further discussed.

\begin{figure}
  \centering
   \includegraphics{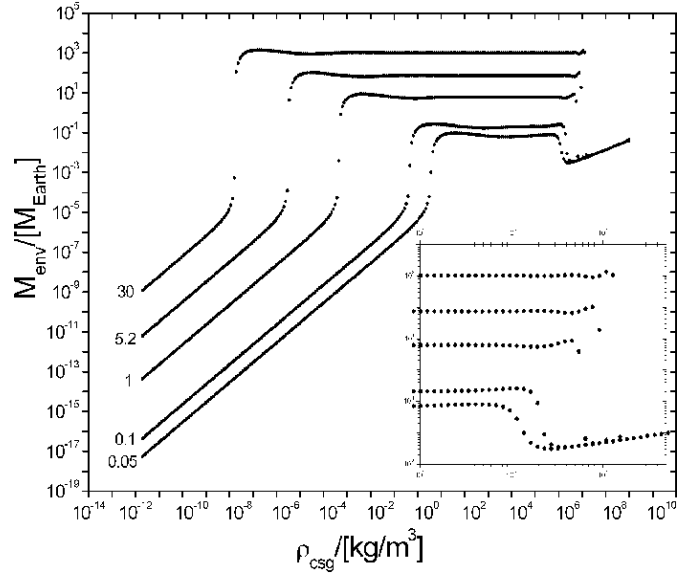}
  \caption {Envelope mass solutions as a function of gas density
            at the core surface, for orbital distances of 0.05, 0.1, 1,
            5.2, and 30 AU.
            Enlargement: The transition from uniform to compact
            envelope solutions is more abrupt
            for protoplanets at large orbital radii. This is a
            consequence of the larger Hill-sphere of outer protoplanets.}
  \label{FigComparManifa}
\end{figure}

\subsubsection{Orbital Distance and Star Class}
\label{SectOrbDist} Orbital distance, together with the masses of
the protoplanet and the parent star determine the protoplanet's
gravitational sphere of influence, the so called Hill-sphere.
Since the available volume for the protoplanet's envelope scales
with the cube of the orbital distance (see
Eq.~(\ref{r_Hill_gen})), the strength of the envelope's
self-gravitating effect depends critically on the distance from
the core to the parent star (see~Fig.~\ref{FigComparManifa}).
Therefore, for the inner protoplanets to have (at least partly)
self-gravitating envelopes, the gas density at the core surface
must be larger than for the corresponding
outer protoplanets.\\
For solutions with compact envelopes (right side of Fig.~\ref{FigComparManifa} and enlargement) the orbital
distance has no impact on the envelope mass, since the radii of the compact inner part are typically several
orders of magnitude smaller than their respective Hill-spheres.\\
The transition from a uniform self-gravitating to a compact
envelope is characterized by a considerable drop in the gas
density for the outer envelope stratifications. In addition,
protoplanets close to the parent star have relatively small
Hill-radii and most of the envelope mass can be found in the
proximity of the core. Therefore, the transition from uniform to
compact envelope for protoplanets close to the parent star is less
abrupt than for more
distant protoplanets, as can be seen in the enlargement of Fig.~\ref{FigComparManifa}.\\
Varying the mass of the parent star is equivalent to changing the
orbital distance of the protoplanet, provided that the gas
temperature stays the same. It follows from Eq.~(\ref{r_Hill_gen})
that $\delta a^{-3} = \delta M_\star$, e.g. changing the orbital
distance of the protoplanet from 5.2 AU to 1 AU is equivalent to
changing the mass of the parent star from $M_\star =
0.21\,M_\odot$ to $M_\star = 30\,M_\odot$. It remains to be seen
whether this equivalence will hold for a more complex model,
because the nebula properties will likely change in accordance
with the known mass-luminosity relation as $M_\star$ is varied.

\subsection{Static Critical Core Mass}
\label{SectStatCCM}

\begin{figure}
  \centering
   \includegraphics[scale=1]{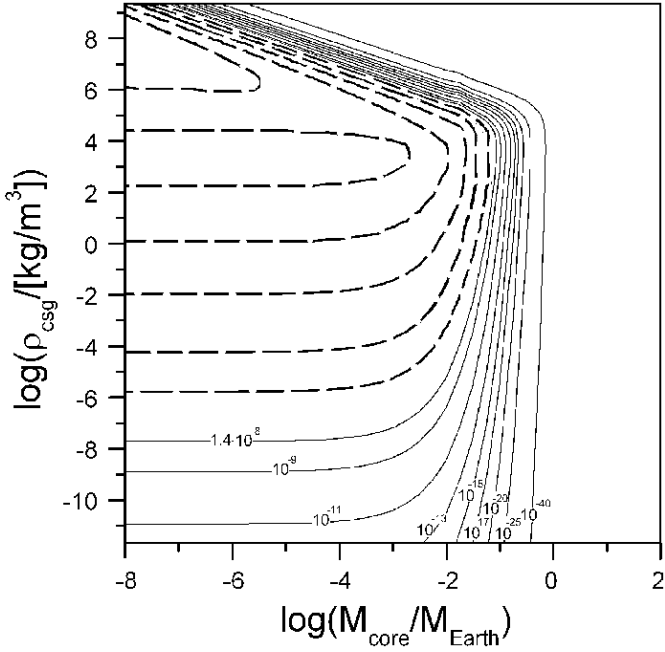}
  \caption{Solution branches - isobars for
  $\varrho_{\rm env}(r_{\rm Hill})=\varrho_{\rm{out}} $ - for
  ($a=5.2$ AU and $T=123$ K) manifold: the standard solar nebula solution branch is
  represented by the innermost solid line; an enhanced nebula with
  $\varrho_{\rm{out}}=10^{-6} \rm {kg \, m^{-3}}$
  (dashed lines) has multiple solution branches; each solution branch has its own maximum core mass,
  and hence local critical mass}
  \label{NebulaeSols}
\end{figure}

There are several definitions of the critical core mass currently
in use. The critical core mass concept has been introduced by
various investigators (e.g. Perri \& Cameron (\cite{perri}),
Mizuno et al. (\cite{miz78}), Mizuno (\cite{mizuno}), Bodenheimer
\& Pollack (\cite{boden86}), Wuchterl (\cite{wuc91a})). As a
starting point, we choose here a definition suggested by Wuchterl
(\cite{wuc91a}), for 'static critical core mass': \emph{No more
static core-envelope
models with increasing core mass exist at the critical mass}.\\
This definition is valid along a (time) sequence of protoplanetary models with increasing $M_{\rm core}$. It is
only along such a sequence, in the context of the static models, that a time evolution with growing cores can
proceed. Essentially, the static critical core mass is the largest core mass for a static protoplanet that can be
embedded in a given nebula, characterized by a nebula gas density, a temperature, and
a distance from a parent star.\\
For the ($a=5.2$ AU and $T=123$ K) manifold this means that, among
the solutions with $\varrho _{\rm env}(r_{\rm Hill})=1.4 \;
10^{-8}\; \rm {kg\,m^{-3}}$ (defined for the minimum mass solar
nebula, e.g. Hayashi et al. \cite{hayashi}), the solution with the
largest core mass determines the static critical mass
(Fig.~\ref{NebulaeSols}, the innermost solid line). This gives a
static critical core mass of
$M_{\rm core,crit}^{\rm{MMSN}}=0.0948M_{\oplus}$.\\
Figure~\ref{NebulaeSols} shows that the value for the critical
core mass exhibits a generally weak dependence on the density of
the surrounding nebula, so the choice of
$\varrho_{\rm{out}}^{\rm{MMSN}}$ from different nebula models is
not critical. For the very dense nebulae (around $10^{-6}\; \rm
{kg\,m^{-3}}$) and depending on the choice of the solution branch
(cf. Sect.~\ref{local_ccm}), the values for the local critical
core masses can span several
orders of magnitude even for the same nebula.\\
The critical core masses for different manifolds are presented in
Table~\ref{TabManifolds}, and are found to depend on the
parameters that affect the hydrostatic balance (cf.
Sect.~\ref{SectEnviro}).

By comparing Figs.~\ref{FigMassManifold} and \ref{FigManifoldReg}
it follows that the natural choice for the \emph{global static
critical core mass,} one which is valid for the whole manifold,
should be the core of the protoplanet which is at the interface of
all four manifold regions (cf. Fig.~\ref{FigManifoldReg}). The
model at the interface has a minimum in the envelope mass, for a
manifold cut along the constant $\varrho_{\mathrm {csg}}$ value.
The interface is also an inflection point for a manifold cut at a
constant $M_{\mathrm {core}}$. The conditions for the \emph{global
static critical core mass} thus are:

\begin {equation} \label{Eqn_CCM}
  \begin{tabular}{llll}
  $\frac{\partial M_{\mathrm {env}}}{\partial M_{\mathrm {core}}}=0$
  &
  &
  &
  $\frac{\partial^{2} M_{\mathrm {env}}}{\partial M_{\mathrm {core}}^{2}}>0$
  \\
  \noalign{\smallskip}
  \noalign{\smallskip}
  \noalign{\smallskip}
  $\frac{\partial M_{\mathrm {env}}}{\partial \varrho_{\mathrm {csg}}}=0$
  &
  &
  &
  $\frac{\partial^{2} M_{\mathrm {env}}}{\partial \varrho_{\mathrm {csg}}^{2}}=0$
  \\
  \end{tabular}
\end{equation}

Since the numerical values for the global critical core masses are
very close to the values of the critical core masses from the
definition suggested by Wuchterl (\cite{wuc91a}), we do not
present the global numerical values separately.

 The values obtained for critical core masses in this model agree well with those of Sasaki
(\cite{sasaki}), who used a similar set of assumptions. However,
such isothermal values are significantly smaller than today's
commonly accepted critical mass values, obtained with the
inclusion of detailed energy transfer, which are typically between
7 and 15 $M_{\oplus}$. The reasons for this are two-fold. Firstly,
we use the equation of state for an ideal gas. Secondly, the
temperature of the isothermal gas is taken from nebula models,
hence the nebula temperature is the temperature of the entire
protoplanet. This is certainly a lower limit for a realistic
temperature of the interior of the protoplanet. Larger critical
core mass values are obtained if the gas temperature is in the
range of the temperatures for the interior of gas giants modelled
with detailed energy transfer (cf. Sect.~\ref{SectTemp}). Clearly,
the correct determination of the critical core mass requires
temperature structure, but the emphasis in this work was not on
quantitative details, but rather on global qualitative features.


\subsubsection{Local Critical Core Mass}
\label{local_ccm}

\begin{figure}
  \centering
   \includegraphics[scale=1]{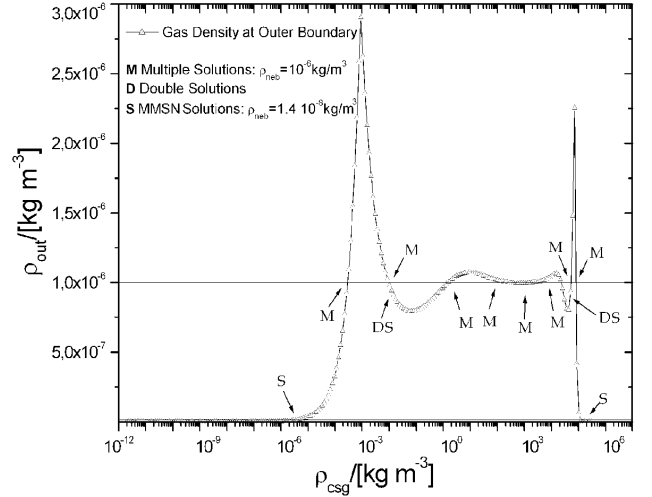}
  \caption{For nebula density enhanced relative to a minimum-mass solar nebula, even
more than two hydrostatic equilibria could exist; $M$:
protoplanetary solutions with $\log M
_{\rm{core}}/[M_{\rm{Earth}}]=-2$ that fit into $\varrho
_{\rm{out}}=10^{-6}\, \rm {kg \, m^{-3}}$ nebula; $DS$: double
solutions, a special case of multiple solutions, cf.
Figs.~\ref{FigRho_out_compar_zoom} and \ref{Double}; $S$:
protoplanetary solutions with the same core, whose envelope fits
into the minimum-mass solar nebula.}
  \label{MultStatOneNeb2}
\end{figure}

From Figs. \ref{NebulaeSols} and \ref{MultStatOneNeb2} one can see
that, for each subcritical core immersed in a nebula, there are at
least two permitted solutions. However, if one considers only the
time-sequence of hydrostatic models with a growing core, it is
clear that solutions with higher density at the core surface
cannot be reached.

\begin{figure}
  \centering
   \includegraphics[scale=1.15]{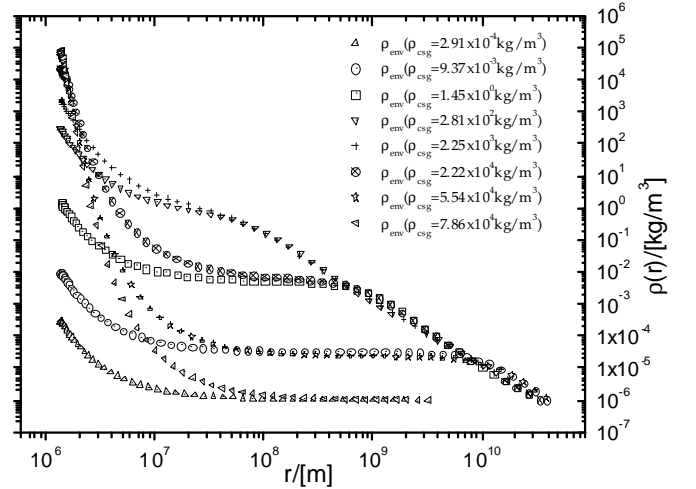}
  \caption{Density profiles for the solutions which fit into the same ($10^{-6}$ kg\,m$^{-3}$) nebula. These
  solutions are labelled with $M$ in Fig.~\ref{MultStatOneNeb2}.}
  \label{rho_multi}
\end{figure}

The situation is more complicated if the protoplanet is embedded
in a denser protoplanetary nebula. Our model clearly predicts
multiple solutions for certain sets of parameters
(Fig.~\ref{MultStatOneNeb2}). Instead of one solution branch for a
given nebula (with two solutions for each core, as for a minimum
mass solar nebula), several solution-branches are possible, again
each with two solutions for a specific core
(Fig.~\ref{NebulaeSols}, dashed solution branches for $\varrho
_{\rm out}=10^{-6}\; \rm kg\,m^{-3}$). Multiple solution-branches
are enabled by envelope self-gravity (cf. Fig.~\ref{rho_multi})
and are due to tidal restrictions imposed by the parent star via
$r_{\rm Hill}$ (cf. Fig. \ref{FigSelf_grav}, region IV).

Each solution branch has one critical core mass, beyond which
there is no static solution, for a sequence of hydrostatic models
with increasing core mass. For the minimum mass solar nebula this
means one critical core mass, according to the definition
suggested by Wuchterl (\cite{wuc91a}). For some denser nebulae,
however, the existence of several branches implies several
\emph{local} critical core masses, where solutions beyond the
critical core mass of the branch are unavailable \textit{locally}.
After reaching the local critical core mass, the planet could, in
principle, continue evolving by 'jumping' to another branch. One
of us has observed similar behavior for certain sets of initial
parameters in hydrodynamical models. The local critical core mass
satisfies the above definition but not Eq.~(\ref{Eqn_CCM}) for the
global critical core mass.

\subsubsection{Double Maxima}
\label{SectDoubleMax}

\begin{figure}
  \centering
   \includegraphics[scale=1]{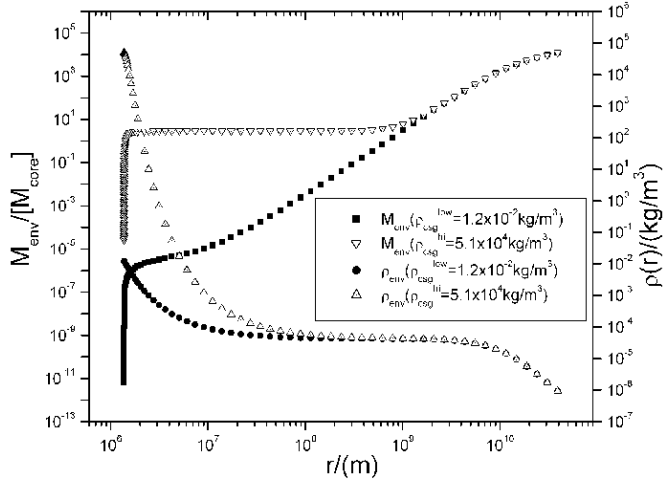}
  \caption{Mass and density radial structure of the special case
  of multiple solutions, where two protoplanets have the same core, almost the same envelope mass,
connect to the same nebula, but have different radial structure.
These
  solutions are labelled $DS$ in Fig.~\ref{MultStatOneNeb2}.}
  \label{Double}
\end{figure}

A special case of multiple solutions can be seen in
Figs.~\ref{FigRho_out_compar_zoom}, \ref{MultStatOneNeb2}, and
\ref{Double} as double peaks in the envelope mass. For every
(subcritical) core, two special solutions, which fit into the same
nebula cloud (i.e. have the same $\varrho(r_{\rm {Hill}}))$ and
have almost exactly the same envelope mass (equal to one part in
$10^4$, or better), are found to exist. Usually these two
solutions have a very similar stratification in the outer parts of
the protoplanet's envelope, but deep inside the protoplanet their
radial structure is quite different (cf.
Sect.~\ref{CorelessVsCore}).

Supercritical cores do not posses such a feature, because the
density profile \emph{always} effectively goes to zero long before
the Hill radius is reached. Therefore there is no significant
contribution to the envelope mass in the outer stratifications,
and the envelope mass increases monotonically with the gas density
at the core surface (cf. Fig.~\ref{FigSelf_grav}).

Envelopes with lower gas density at the core surface,
$\varrho_{\rm {csg}}^{\rm{low}}$, (Fig.~\ref{Double}) have a
maximum possible mass (for the corresponding manifold) because the
envelope gas density at the core surface is low enough to ensure
uniformity for the major part of the radial structure.
Consequently, the envelope density does not substantially decrease
from the core-surface value. At the same time, $\varrho_{\rm
{csg}}$ is high enough to allow significant mass contributions
from the outer parts of the envelope, where the volume (and
therefore the mass, for a given density) per unit radius, is the
largest. Values for such maximum envelope masses are
tabulated in Table~\ref{TabManifolds}.\\
Envelopes with higher $\varrho_{\rm {csg}}^{\rm{high}}$ build up
the self-gravitating effect (SG effect starts as soon as
$M_{\rm{env}}\approx M_{\rm{core}}$) very close to their core.
Because of the very strong self-gravitating effect
($M_{\rm{env}}\approx 3M_{\rm{core}}$ for the innermost regions),
the
radial density fall-off close to the core (Fig.~\ref{Double}) is strong.\\
A new, \emph{effective core} is formed from the dense
envelope-gas-layer wrapped tightly around the core. In this case,
the envelope density distribution resembles one with the core (and
the radius) of this effective core. In stratifications where the
envelope mass becomes comparable to the effective core, another
self-gravitating effect changes the radial envelope density
distribution to
$\varrho_{\rm{env}}\propto r^{-2}$.\\
For a particular choice of $\varrho_{\rm {csg}}^{\rm{high}}$, the
envelope density profile in the outer stratifications matches that
of $\varrho_{\rm {csg}}^{\rm{low}}$, thus making the mass of both
envelopes almost equal.

\subsection{Temperature-Mass Invariance}
\label{SectT-M}

It has been noted that, if mass and distance are measured in a
system of appropriate units (i.e. mass in units of core mass, and
distance in units of core radii), solution manifolds with
different temperatures are almost identical, except for a shifting
on a core-mass-axis, according to the relation:

    \begin{equation}\label{temp_mass_prop}
      \frac{T_1}{T_2} = (\frac{M_1}{M_2})^{2/3}
    \end{equation}
    that can be derived for homologous envelopes satisfying
$\varrho_1(r_1/r_{\rm core,1})=\varrho_2(r_2/r_{\rm core,2})$, for
any pair of $r_1$ and $r_2$ such that $r_1/r_{\rm
core,1}=r_2/r_{\rm core,2}$. In other words, the radial profile of
a certain protoplanet with core mass $M_1$ and temperature $T_1$
will be the same as the radial profile of another protoplanet with
core mass $M_2$ and temperature $T_2$, if
Eq.~(\ref{temp_mass_prop}) is obeyed, and if the mass is measured
in units of core mass and the length in units of core radius.\\
This is true for all manifold regions, sub- and super-critical,
self-gravitating or not. Note that in Fig.~\ref{FigComparManifT}
the non-self-gravitating region was not affected by a change in
envelope temperature, but relation (\ref{temp_mass_prop}) does
hold even for non-self-gravitating envelopes, since it connects
envelopes with different temperatures \emph{and} core masses.
Fig.~\ref{FigComparManifT} was plotted for different temperatures,
but constant core mass.

\section{Discussion and Conclusions}
\label{SectDisc}

In an effort to obtain a global overview of hydrostatic protoplanetary equilibria, we have chosen a simple
physical model so as to be able to clearly understand the interaction of competing processes.

Our use of relatively simple physics has several consequences;
because the ideal gas equation of state is used, gas particles are
`soft', and can be compressed as much as is needed, in effect
overestimating the importance of gravity relative to gas pressure,
when large envelope-gas-pressure is applied. A comparison of the
ideal gas EOS to the numerical Saumon-Chabrier EOS shows
disagreement for the $\log \, T=2.1$ isotherm and densities above
$\varrho=40\; \rm{kg \, m^{-3}}$. This would indicate that the
non-ideal EOS is needed for high-density effective cores.

It has been noted that manifold properties are insensitive to
variation of orbital distance $a$ or mass of the parent star
$M_\star$, as long as $a \, M_\star^{-3}=const$ holds (cf.
Sect.~\ref{SectOrbDist}). Also, solutions whose envelope
temperature and core mass obey relation (\ref{temp_mass_prop}) are
found to be the same, if appropriate units for mass (i.e. core
mass) and length (i.e. core radius) are used. This indicates the
existence of analytic solutions for some envelope regimes, through
certain dimensionless scaling variables. Such a treatment is,
however, out of the scope of the present paper.

An envelope gas temperature is equal to the nebula $T$ throughout
the protoplanet, and that certainly underestimates the thermal
pressure and hence reduces the values for the critical core mass.
However, from Equ.~\ref{temp_mass_prop}, one can show that for a
more realistic estimate of the envelope temperature representative
for the young planets (5000 K) critical core mass values are
overestimated ($\sim\,24.5 M_\oplus$), because of envelope
isothermality/lack of an energy transport equation and use of
ideal-gas EOS, when compared to canonical critical core mass
values from protoplanetary models with detailed microphysics.

Both the local and the global critical core masses signal the end of the availability of the hydrostatic
solutions. In the case of the local critical core mass, non-availability holds for a small region of the parameter
space around the local critical core mass, while for the global critical core mass this is true for every core
larger than the critical core mass. The significant difference between the two types of critical core mass is
that, at the global critical core mass (and above), the non-isothermal effects are crucial in shaping the
structure of the protoplanetary envelopes, and are present throughout the parameter space. These non-isothermal
effects are important for determining the details of the dynamical disk-planet interaction.

The critical core mass values obtained in this model are almost
two orders of magnitude smaller than the canonical critical core
masses which incorporate detailed energy transfer. Thus, if
subcritical or just-critical regimes of a dynamical disk-planet
interaction are to be investigated through a model that is locally
isothermal, the planet mass should be set appropriately. Most of
the present locally-isothermal disk-planet models (e.g. Kley
\cite{kley}, D'Angelo et al. \cite{dangelo02}, \cite{dangelo03},
Nelson \& Papaloizou \cite{nelson04}) operate with planets which
\emph{should be deep in the super-critical regime.}

A solution set from our model encompasses solutions that are
reminiscent of the planets in the various stages of evolution
(from small rocks embedded in the dilute nebula to the mature
planets as we know them), and of various configurations (the
telluric planets of region I in Fig.~\ref{FigManifoldReg}, and the
gas giants of region II). The 'nebula' and 'mature planet' regimes
are the physically intuitive beginning and end phase of planetary
evolution. However, the 'protoplanet' regime presents us with an
interesting region in parameter space, where planets could make
the transition from 'infancy' to 'maturity'. Depending on the
detailed structure and the dynamics of the surrounding nebula, it
is easy to conceive a standard scenario of planet formation. That
is, the accretion of nebula gas onto a supercritical protoplanet.
Other scenarios could be imagined as well, e.g. a massive
protoplanet could release a major part of its envelope to reach
the appropriate equilibrium, or it could dramatically condense its
otherwise mostly gaseous envelope. Amounts of dust in the
environment will doubtless play a very important role in the
process.

In conclusion, several important features of the solution set have to be mentioned:
\begin{enumerate}
    \item Two basic types of the envelope equilibria are found for protoplanets:
        \subitem $\bullet$ \emph{uniform}; the density of the envelope gas
                drops weakly from the core to the outer boundary
        \subitem $\bullet$ \emph{compact}; a dense gas layer forming an effective core,
                and a very low, exponentially decreasing, gas density further
                out\\
                Both types can be self-gravitating or
                non-self-gravitating, dividing the solution manifold into four distinct regions.
    \item As a consequence of the envelope's self-gravitating effect,
    a wide range of possible envelope solutions exists.
    \item We have developed a new concept for the global static critical core mass, which marks
    the contact point of all four qualitatively different types of protoplanets. This concept is based on a qualitative
    change of the envelope properties while considering a complete set of available solutions (a solution manifold),
    as opposed to the critical core mass definitions which are valid only for a solution
    subset fitting a particular nebula.
    \item For every subcritical core there are at least two
    envelope solutions possible (a self-gravitating one and a
    non-self-gravitating one) for a given nebula, and for a
    certain nebula parameters the number of the possible envelope
    solutions can be even larger. Such nebulae also have
    multiple (local) critical core masses.
    \item The global static critical core mass value
     is shown to decrease with the increasing orbital distance $a$, mainly
     because of the decrease in the temperature of the surrounding nebula.

\end{enumerate}



\begin{thebibliography}{}

  \bibitem[1986]{boden86} Bodenheimer, P., \& Pollack, J. B. 1986,
  Icarus, 67, 391-408

  \bibitem[2000]{bodenheimer} Bodenheimer, P., Hubickyj, O., \& Lissauer, J. J. 2000, Icarus,
143, 2

  \bibitem[1969]{carnahan-starling} Carnahan, N. F \& Starling, K. E. 1969, J. Chem. Phys. 51, 635-636

  \bibitem[2002]{dangelo02}D'Angelo, G., T. Henning and W. Kley 2002, Astron. Astroph. 385, 647

  \bibitem[2003]{dangelo03} D'Angelo, G., W. Kley and T. Henning 2003, Astroph. Journ. 586,
540

  \bibitem[2001]{garvan} Garvan, F. 2001, The Maple Book (Chapman \&
  Hall)

  \bibitem[1999]{guillot} Guillot, T. 1999, Science, 286, 72-77

  \bibitem[1985]{hayashi} Hayashi, C., Nakazawa, K., \& Nakagawa, Y. 1985, Protostars and
  Planets II, 1100

  \bibitem[2001]{ikoma} Ikoma, M., Emori, H., \& Nakazawa, K. 2001, Aph. J., 553, 999-1005

  \bibitem[1991]{kippenhahn} Kippenhahn, R. \&
  Weigert, A. 1991, Stellar Structure and Evolution
  (Springer-Verlag)

  \bibitem[1999]{kley} Kley, W. 1999, MNRAS, 303, 696

  \bibitem[1970] {kusaka} Kusaka, T., Nakano, T., \& Hayashi, C.
  1970, Prog. Theor. Phys., 44, 1580-1596

  \bibitem[1978] {miz78} Mizuno, H., Nakazawa K., \& Hayashi, C.
  1978, Prog. Theor. Phys., 60, 699-710

  \bibitem[1980] {mizuno} Mizuno, H. 1980, Prog. Theor. Phys., 64, 544-557

  \bibitem[2004] {nelson04} Nelson, R. P. \& Papaloizou, J. C. B. 2004, MNRAS, 350-3, 849-864

  \bibitem[1974]{perri} Perri, F., \& Cameron, A. G. W. 1974, Icarus 22, 416-425

  \bibitem[1989]{sasaki} Sasaki, S. 1989, Astron. Astrophys., 215, 177-180

  \bibitem[1992]{shu}Shu, F. H. 1992, The Physics of Astrophysics, Volume II,
  University Science Books

  \bibitem[1995]{SCvH} Saumon, D., Chabrier, G., \& van Horn, H.~M.
  1995, ApJS, 99, 713

  \bibitem[1995]{stahl} Stahl, B., Kiessling, M. K-H., \& Schindler, K.
  1995, Planet. Space Sci., 43, 271-282

  \bibitem[1982]{stevenson} Stevenson, D. J. 1982, Planet. Space Sci.,
  Vol. 30, No. 8, 755-764

  \bibitem[1957]{schwarzschild} Schwarzschild, M. 1957, Structure and Evolution of the Stars,
  Princeton University Press

  \bibitem[1991a]{wuc91a} Wuchterl, G. 1991a, Icarus, 91, 39-52

  \bibitem[1991b]{wuc91b} Wuchterl, G. 1991b, Icarus, 91, 53-64

  \bibitem[1993]{wuc93} Wuchterl, G. 1993, Icarus, 106, 323-334

  \bibitem[2000]{ppiv} Wuchterl, G., Guillot, T., \& Lissauer, J.
  J. 2000, Protostars and Planets IV, 1081


\end{thebibliography}
\end{document}